\documentstyle[12pt]{article}
\newcommand{\tline}{\setlength{\baselineskip}{0.84cm}}

\begin{document}
\tline
\tline
\begin {center}
{\large {\bf On the Original Proof by {\it Reductio ad Absurdum}}}\\
{\large {\bf of the Hohenberg-Kohn Theorem}} \\
{\large {\bf for Many-Electron Coulomb Systems}}
\end{center}
\vspace{1cm}
\begin{center} 
Eugene S. KRYACHKO\footnote[1]{FAX: +32 (4) 366 3413; E-mail: eugene.kryachko@ulg.ac.be}
\end{center}
\vspace{1cm}
{\normalsize \centerline{Bogoliubov Institute for Theoretical Physics, Kiev, 03143 Ukraine}}
{\normalsize \centerline{and}}
{\normalsize \centerline{Department of Chemistry, Bat. B6c, University of Liege,}}
{\normalsize \centerline{Sart-Tilman, B-4000 Liege 1, Belgium}}
\vspace{3cm}
\begin{abstract}
\vspace{0.5cm}
It is shown that, for isolated many-electron Coulomb systems with Coulombic external potentials, the usual {\em reductio ad absurdum} proof of the Hohenberg-Kohn theorem is unsatisfactory since the to-be-refuted assumption made about the one-electron densities and the assumption about the external potentials are not compatible with the Kato cusp condition. The theorem is, however, provable by more sophisticated means and it is shown here that the Kato cusp condition actually leads to a satisfactory proof. 
\end{abstract}
\vspace{1cm}

\newpage 

\centerline{\bf {1. Introduction}}
\vspace{0.25cm}

The Hohenberg-Kohn theorem [1] underlies the foundation of the density functional theory [2] and since 1964 when it was formulated and proved by {\em reductio ad absurdum}, it has significantly influenced the state of art of quantum theory of atoms, molecules, clusters, and solids. 

The aim of the present work is an attempt to rethink and reanalyze the original proof by {\em reductio ad absurdum} of the Hohenberg-Kohn theorem. It is shown here that, although the result is a generally correct one, the original proof cannot be maintained when the external potential is of Coulomb form because it implies that the supposed wave functions must violate the Kato electron-nuclear cusp conditions. However, more sophisticated proofs can be adduced, in particular directly from the Kato theorem, which avoid this problem.
\vspace{1cm}

\centerline{\bf {2. General Part}}
\vspace{0.25cm}

On p. B864 of their work [1], Hohenberg and Kohn state that they ``... develop an exact formal variational principle for the ground-state energy, in which the density" $\rho({\bf r})$ (in a widely accepted notation) ``is the variable function. Into this principle enters a universal functional" $F[\rho({\bf r})$], ``which applies to all electronic systems in their ground state no matter what the external potential is." Following Hohenberg and Kohn [1], let us consider ``a collection of an arbitrary number of electrons, enclosed in a large box and moving under the influence of an external potential 
$v({\bf r})$" where ${\bf r} \in \Re^3$ ``and mutual Coulomb repulsion." The Hamiltonian $H_v^N$ of a given $N$-electron system casts as 
\begin{equation}
H_v^N = T_e^N + V_{ee}^N + V^N
\label{1}
\end{equation}
where $T_e^N$ is the kinetic energy operator of $N$ electrons, $V_{ee}^N$ is the corresponding interelectronic Coulomb operator, and 
\begin{equation}
V^N = \Sigma_{i=1}^N v({\bf r}_i)
\label{2}
\end{equation}
is the total external potential. Hohenberg and Kohn [1] further assume (p. B865) that $H_v^N$ possesses the least bound-state (ground-state) wavefunction $\Psi_o({\bf r}_1, {\bf r}_2, ..., {\bf r}_N) \in {\cal H}^1(\Re^{3N})$ (spins are omitted for simplicity) and the latter is nondegenerate within the total coordinate - spin representation. ${\cal H}^1(\Re^{3N})$ is the Sobolev space of $N$-electron wavefunctions whose norms defined as $(\int \prod_{i=1}^N d^3{\bf r}_i (\mid \Psi_o \mid^2 + \mid \nabla \Psi_o \mid^2 )^{1/2} < \infty$. Let us then define the corresponding ground-state one-electron density [3] 
\begin{equation}
\rho_o({\bf r}) \equiv N \int \prod_{i=2}^N d^3{\bf r}_i \mid\Psi_o({\bf r}, {\bf r}_2, ..., {\bf r}_N)\mid^2 \in D_o^N,
\label{3}
\end{equation}
where $D_o^N \equiv \{ \rho_o \in L^1(\Re^3) \mid \rho_o \geq 0, \sqrt{\rho_o} \in {\cal H}^1(\Re^3), \int d^3{\bf r} \rho_o({\bf r}) = N \}$, ``which is clearly a functional of $v({\bf r})$" (p. B865, Ref. [1]) since $H_v^N$ is explicitly determined by $v({\bf r})$ under the fixed $N, T_e^N$, and $V_{ee}^N$. That is, there exist such mappings
\begin{equation}
v({\bf r}) \stackrel{{\cal C}_N}{\Rightarrow} \Psi_o({\bf r}_1, {\bf r}_2, ..., {\bf r}_N) \stackrel{{\cal D}_N}{\Rightarrow} \rho_o({\bf r})
\label{4}
\end{equation}
from $L^{3/2}(\Re^3) + L^\infty(\Re^3)$ as the domain of external potentials [4, 5] (see precisely Eq. (2.4) in Ref. [6]) to ${\cal H}^1(\Re^{3N})$ and further to $D_o^N$ if $H_v^N$ does possess the nondegenerate ground state. Notice that belonging of any $v({\bf r})$ to $L^{3/2}(\Re^3) + L^\infty(\Re^3)$ does not guarantee that the corresponding $H_v^N$ of the type (1) has this property [4]. Notice also that any Coulomb potential, $v({\bf r}) = -Z_\alpha/\mid{\bf r} - {\bf R}_\alpha\mid$, or any finite linear combination of Coulomb potentials, $v({\bf r}) = -\Sigma_{\alpha=1}^M Z_\alpha/\mid{\bf r} - {\bf R}_\alpha\mid$, belong to $L^{3/2}(\Re^3) + L^\infty(\Re^3)$ [5]. The mapping ${\cal C}_N$ in (4) is defined as that from $L^{3/2}(\Re^3) + L^\infty(\Re^3)$ as the subdomain of external potentials for which $H_v^N$ possesses the ground state into ${\cal H}^1(\Re^{3N})$. It is readily to prove that ${\cal C}_N$ is invertible (see, e. g., Ref. [5]) that is any pair of external potentials differed from each other by more than a constant determine a pair of different ground-state wavefunctions. The other mapping in (4) is the mapping ${\cal D}_N$ from the ground-state $N$-electron eigenwave functions onto the set $\tilde{D}_o^N \subset D_o^N$ of the ground-state one-electron densities. An invertibility of the mapping ${\cal D}_N$ relies on the Hohenberg-Kohn theorem (see also Eqs. (27)-(29) in Ref. [5]). 

{\bf Hohenberg-Kohn theorem [1]: ``$\mathbf{\em{v}({\bf r})}$ is a unique functional" of $\mathbf{\rho({\bf r})}$, ``apart from a trivial additive constant."}

{\bf Proof} (p. B865, Ref. [1]): ``The proof proceeds by {\it reductio ad absurdum}." We {\em assume} the existence of two ``external" potentials $v_1({\bf r})$ and $v_2({\bf r})$ such that
\begin{equation}
v_1({\bf r}) \neq v_2({\bf r}) + \mbox{constant}.
\label{5}
\end{equation}
Via Eqs. (2) and (1), $v_1({\bf r})$ and $v_2({\bf r})$ define the Hamiltonians $H_1^N$ and $H_2^N$ associated with two different $N$-electron systems. Let us further {\em assume} the existence of the ground-state normalized wavefunctions $\Psi_o^{(1)} \in {\cal H}^1(\Re^{3N})$ and $\Psi_o^{(2)} \in {\cal H}^1(\Re^{3N})$ of $H_1^N$ and $H_2^N$, respectively. By virtue of Eq. (3), $\Psi_o^{(1)}$ and $\Psi_o^{(2)}$ yield the corresponding ground-state one-electron densities $\rho_o^{(1)}({\bf r})$ and $\rho_o^{(2)}({\bf r})$. Hohenberg and Kohn [1] finally {\em assume} that

{\bf (i)} $\Psi_o^{(1)} \neq \Psi_o^{(2)}$

{\bf (ii)} $\rho_o^{(1)}({\bf r}) = \rho_o^{(2)}({\bf r}) = \rho_o({\bf r})$.

Applying the Rayleigh-Ritz variational principle, one obtains
\begin{eqnarray}
E_o^{(1)} &=& \langle \Psi_o^{(1)}\mid H_1^N\mid \Psi_o^{(1)} \rangle 
\stackrel{\bf (i)}{<} \langle \Psi_o^{(2)}\mid H_1^N\mid \Psi_o^{(2)} \rangle \nonumber \\
&\stackrel{Eq.(5)}{=}& \langle \Psi_o^{(2)}\mid H_2^N\mid \Psi_o^{(2)} \rangle + \langle \Psi_o^{(2)}\mid V_1^N - V_2^N \mid \Psi_o^{(2)} \rangle \nonumber \\ 
&=& E_o^{(2)} + \int d^3 {\bf r} [v_1({\bf r}) - v_2({\bf r})] \rho_o({\bf r})
\label{6}
\end{eqnarray}
and 
\begin{eqnarray}
E_o^{(2)} &=& \langle \Psi_o^{(2)}\mid H_2^N\mid \Psi_o^{(2)} \rangle 
\stackrel{\bf (i)}{<} \langle \Psi_o^{(1)}\mid H_2^N\mid \Psi_o^{(1)} \rangle \nonumber \\
&\stackrel{Eq.(5)}{=}& \langle \Psi_o^{(1)}\mid H_1^N\mid \Psi_o^{(1)} \rangle + \langle \Psi_o^{(1)}\mid V_2^N - V_1^N \mid \Psi_o^{(1)} \rangle \nonumber \\ 
&=& E_o^{(1)} - \int d^3 {\bf r} [v_1({\bf r}) - v_2({\bf r})] \rho_o({\bf r})
\label{7}
\end{eqnarray}
where the used formulas are indicated above the signs. 

Hohenberg and Kohn then conclude (p. B865, Ref. [1]) that adding (6) 
to (7) ``leads to the inconsistency"
\begin{equation}
E_o^{(1)} + E_o^{(2)} < E_o^{(1)} + E_o^{(2)}, 
\label{8}
\end{equation}
and therefore, (8) implies that the assumption {\bf (ii)} fails. ``Thus $v({\bf r})$ is (to within a constant) a unique functional of" $\rho({\bf r})$, ``since, in turn, $v({\bf r})$ fixes" $H_v^N$ ``we see that the full many-particle ground state is unique functional of" $\rho({\bf r})$. Q. E. D. 

Examine Eq. (8). It is obviously self-contradictory. (8) is deduced under the assumption that (5) is true together with the to-be-refuted assumptions {\bf (i)} and {\bf (ii)}, both composing the negation of the statement of the Hohenberg-Kohn theorem. (8) then appears to be absurd in a sense of being obviously false and therefore the statement of the Hohenberg-Kohn theorem is correct. Logically speaking, the fact that Eq. (8) looks absurd implies that, first, one of the to-be-refuted assumptions, {\bf (i)} or {\bf (ii)}, or simultaneously both, {\bf (i)} and {\bf (ii)}, lead to the contradiction with (5) {\em or}, second, they are a priori invalid in a sense that one of them or both are somehow incompatible with (5). In the latter case, the statement of the Hohenberg-Kohn theorem is invalid unless it is proved in the other 
way. Note also that {\bf (i)} and {\bf (ii)} are obviously inconsistent for $N = 1$ and for the two-electron noninteracting systems. Explicitly, after deriving (8), one has to consider the following cases:

{\bf (I)} $\Psi_o^{(1)} = \Psi_o^{(2)} = \Psi_o$. 

This one directly yields $\rho_o^{(1)} = \rho_o^{(2)} = \rho_o$, that is, {\bf (ii)} does hold. It also implies that
\begin{equation}
V_1^N \equiv V_2^N \equiv E_o - \frac{(T_e^N + V_{ee}^N) \Psi_o}{\Psi_o}
\label{9}
\end{equation}
if $V_1^N$ and $V_2^N$ are multiplicative operators, as suggested by Eq. (2). It is trivial to conclude that (9) contradicts to (5). However, there is no ``inconsistency" because the last terms in the last lines of Eqs. (6) and (7) simply vanish.  

{\bf (II)} $\Psi_o^{(1)} \neq \Psi_o^{(2)}$ and $\rho_o^{(1)} \neq \rho_o^{(2)}$. 

This case precisely lies in the line of the original arguments by Hohenberg and Kohn [1] proving thus that different external potentials determine different ground-state one-electron densities. 
 
{\bf (III)} $\Psi_o^{(1)} = \Psi_o^{(2)}$ and $\rho_o^{(1)} \neq \rho_o^{(2)}$.

These two relations contradict to each other due to (3). 

{\bf (IV)} A self-contradiction ({\em ad absurdum}) of Eq. (8) might also mean that the to-be-refuted assumptions {\bf (i)} or/and {\bf (ii)} in the original proof of the Hohenberg-Kohn theorem are self-contradictory with Eq. (5) and this is precisely the case of real many-electron Coulomb systems with Coulombic external potentials. In other words, the original {\em reductio ad absurdum} proof of the Hohenberg-Kohn theorem based on the assumption (5) is incompatible with the {\it ad absurdum} assumption {\bf (ii)} due to the validity of the Kato theorem for such systems [6]. A similar statement is valid with regard to the proof by {\em reductio ad absurdum} of the invertibility of the mapping ${\cal D}_N$ in Ref. [5].

Let us recall that the Kato theorem [6] (see also Refs. [7, 3]) determines the character of the singularity of the exact $N$-electron eigenwavefunction of $H_v^N$ at the electron-nucleus coalescences where 
the Coulombic external potential $v({\bf r})$ (see Eq. (2.2) and the conditions i) and ii) on p. 154 and Theorem I on p. 156 of Ref. [6]),
\begin{equation}
v({\bf r}) = -\Sigma_{\alpha=1}^M \frac{Z_\alpha}{\mid{\bf r} - {\bf R}_\alpha\mid}, 
\label{10}
\end{equation}
(in atomic units) is singular. In Eq. (10), the $\alpha$th nucleus with the nuclear charge $Z_\alpha$ is placed at ${\bf R}_\alpha \in \Re^3$. Any $N$-electron eigenwave function $\Psi$ of $H_v^N$ with $v({\bf r})$  of the form (10) and its one-electron density $\rho_\Psi$ then satisfy the electron-nucleus cusp conditions
\begin{eqnarray}
\frac{d}{dr_i} \Psi^{av}({\bf r}_1, {\bf r}_2, ..., {\bf r}_{i-1}, r_i, {\bf r}_{i+1}, ..., {\bf r}_N ) \mid_{r_i = R_\alpha} &=& -Z_\alpha \Psi({\bf r}_1, {\bf r}_2, ..., {\bf r}_{i-1}, {\bf R}_\alpha, {\bf r}_{i+1}, ..., {\bf r}_N ), \nonumber \\ 
\frac{d}{dr} \rho_\Psi^{av}(r) \mid_{r = R_\alpha} &=& -2Z_\alpha \rho_\Psi({\bf R}_\alpha)  
\label{11}
\end{eqnarray}
where the superscript `$av$' means the average of $\Psi$ over an infinitesimally small sphere centered at ${\bf r}_i$ and that of $\rho_\Psi$ at ${\bf r}$, and where $i = 1, 2, ..., N$ in the former relationship. 

Therefore, the true one-electron density of the given $N$-electron system, moving in the Coulombic field of point nuclei, exhibits cusps (local maxima) at the positions of the nuclei. Assuming that the ground-state one-electron density $\rho_o({\bf r})$ is given and analyzing its topology over the whole coordinate space $\Re^3$, one locates the positions of its cusps and evaluates there the lhs of the last equation (11). Altogether, the positions of the electron-nucleus cusps (as being always negative that follows from Eq. (11)) and the halves of the radial logarithmic derivatives of $\rho_o^{av}(r)$, taken with the opposite sign at these points, fully determine the Coulombic external potential $v({\bf r})$, Eq. (10), of the given system. That is, $\rho_o({\bf r})$ uniquely determines the external Coulombic potential $v({\bf r})$. Such interpretation of the Hohenberg-Kohn theorem for Coulombic external potentials was originally proposed by Coleman [8], Bamzai and Deb [9], Smith [10], and E. Bright Wilson (quoted by L\"owdin [11]). Also, this naturally proves an invertibility of the mappings ${\cal C}_N$, ${\cal D}_N$, and ${\cal D}_N{\cal C}_N$. Therefore, if a given pair of $N$-electron systems with the Hamiltonians $H_1^N$ and $H_2^N$ of the type (1) are characterized by the same ground-state one-electron densities (that is equivalent to the to-be-refuted assumption {\bf (ii)}), their external potentials $v_1({\bf r})$ and $v_2({\bf r})$ of the form (10) are identical, and thus $H_1^N$ and H$_2^N$ are identical as well. This contradicts (5) and hence, the assumption {\bf (ii)} cannot be used, rigorously speaking, in the proof via {\it reductio ad absurdum} of the Hohenberg-Kohn theorem together with the assumption (5) for the Coulomb class of external potentials. In other words, (5) and {\bf (ii)} are Kato-type incompatible to each other for purely Coulombic external potentials. 

Vice versa, the nuclei of the given $N$-electron system are isolated 3D point attractors behaving topologically as critical points of rank three and signature minus three [12]. However, there exist some ``specific" many-electron systems whose ground-state one-electron densities have local maxima at non-nuclear positions [13]. These local non-nuclear maxima might be the true ones or appear as a consequence of an incomplete, inadequate quantum mechanical treatment. Therefore, despite the present conclusion that in the original proof by {\it reductio ad absurdum} of the Hohenberg-Kohn theorem, one of its to-be-refuted assumption {\bf (ii)} is incompatible, by virtue of the Kato theorem, with the assumption (5)\footnote[2]{For a similar proof of the ensemble generalization of the Hohenberg-Kohn theorem see Section II of Ref. [14].}, the Kato theorem itself guarantees the existence of the one-to-one correspondence between the Coulomb class of external potentials (10) and the ground-state one-electron densities for nearly all many-electron except probably those ``specific" ones. However, local non-nuclear maxima of the ground-state one-electron densities of these ``specific" many-electron systems can be easily excluded since, first, they are not cusps at all and second, they disappear under some changes (i. e., bond lengthening) of the nuclear skeleton [13]. 

Summarizing, the Kato theorem implicitly contains the proof of the Hohenberg-Kohn theorem for $N$-electron systems with the Coulomb class of external potentials and assures an invertibility of the one-to-one mapping ${\cal D}_N{\cal C}_N$ between the Coulombic external potentials and thus between many-electron Hamiltonian $H_v^N$ and the corresponding nondegenerate ground-state one-electron densities (see also Ref. [5]). Obviously, due to its original formulation [6], the Kato theorem cannot be used for other types of external potentials (that is why the present treatment is only confined to the Coulombic ones of the type (10)) although it might be generalized to include the subclass of the singular ones, like, for example, a Yukawa potential that results in nonvanishing cusps (see p. 156 of Ref. [6] for the definition of the class of generalized Coulomb potentials).

According to the work [1] of Hohenberg and Kohn, the Hohenberg-Kohn theorem implies the existence of the universal energy density functional for any isolated many-electron Coulomb system. This statement has been usually interpreted as the second Hohenberg-Kohn theorem [2]. In the density functional theory, there exist two rigorous constructions of the universal energy density functionals based on their own rigorous proofs of the Hohenberg-Kohn theorem. This is the Levy-Lieb energy density functional [15, 4] and the energy density functional based on the group of the local-scaling deformations in $\Re^3$ consisting of topogical deformations mapping or topologically deforming any pair of one-electron densities to each other [16, 2c]. The related Jacobian of such deformation gives rise to the first-order nonlinear differential equation, a so called ``Jacobian equation" [17, 18] whose solution, namely, the corresponding deformation, does exist within this approach and it is unique [16, 2c]. Solving then the ``Jacobian equation" enables to determine the deformation for any pair of well-behaved one-electron densities and to consistently extend the action of the local-scaling deformation group onto ${\cal H}^1(\Re^{3N})$ [16, 2c]. This larger local-scaling transformation group partitions ${\cal H}^1(\Re^{3N})$ into disjoint classes, orbits (see Refs. [16, 2c] for details). All orbits exhaust ${\cal H}^1(\Re^{3N})$ and within an each given orbit, there establishes the one-to-one correspondence between its wavefunctions and the one-electron-densities. That is, these orbits are endowed with the characteristic that there are no two wavefunctions belonging to the same orbit that have the same density. Each orbit, say ${\cal O}^{[\alpha]}$, is determined by its generator wavefunction $\Psi_g^{[\alpha]}$. Therefore, for a given orbit ${\cal O}^{[\alpha]}$, one defines the energy density functional $E_\alpha[\rho({\bf r})]$ as merely the restriction of the energy functional $E[\Psi]$ on those wavefunctions that belong to the $\alpha$th orbit, $\Psi \in {\cal O}^{[\alpha]}$. It is trivial to prove that, first, there are as many different energy density functionals as the orbits in ${\cal H}^1(\Re^{3N})$, and second, that each density functional $E_\alpha[\rho({\bf r})]$ implicitly depends on the generator wavefunction $\Psi_g^{[\alpha]}$. Evidently, the exact ground-state $N$-electron eigenwave function of the given Hamiltonian operator $H$ belongs to a certain orbit, called the Hohenberg-Kohn one, ${\cal O}^{[HK]}$ [16, 2c]. Within ${\cal O}^{[HK]}$, the Levy-Lieb energy density functional [15] exactly coincides with $E_{HK}[\rho({\bf r})]$ defined within the local-scaling deformation approach [16]. 

The explicit form of any energy density functional $E_\alpha[\rho({\bf r})]$ for any $\alpha$ has been obtained and the corresponding variational Euler-Lagrange equation has been also derived in Ref. [16] (see also Ch. 7 and 8 of Ref. [2c]). The rigorous mathematical framework of the local-scaling deformation approach to the density functional theory based on the ``Jacobian equation" has recently been elaborated in Ref. [17]. The local-scaling deformation approach to the density functional theory has been also generalized 
for the spin densities, the momentum space representation, excited states, fractional occupation numbers, and finally to study nonadiabatic effects. The corresponding analogue of the Kohn-Sham approach has been also formulated in terms of the orbit generators (see Ch. 8 in Ref. [2c]). A variety of theoretical and computational applications of the local-scaling deformation density functional theory to atoms and molecules has recently been elaborated as well (see Ref. [19] and references therein).   
\vspace{1cm}

\centerline{\bf {Acknowledgment}}

I gratefully thank all colleagues with whom I have a great priviledge to work and to discuss the density functional theory, especially and particularly, Erkki Br\"andas, Frank Harris, Jacob Katriel, Ingvar Lindgren, Eduardo Lude\~na, Nimrod Moiseyev, Joe Paldus, Vedene Smith, and Brian Sutcliffe. I would like indeed also to thank Francoise Remacle for warm hospitality and the F.R.F.C. 2.4562.03F (Belgium) for fellowship. I also appreciate the reviewers for their extremely valuable comments and suggestions. 

\end{document}